\documentclass[12pt]{article}

\begin{document}

\title{Solar X-rays as Signature for New Particles}

\author{K.~Zioutas\\ \it \small University of Thessaloniki, Greece $\&$ CERN, Geneva,
Switzerland \\ \small E-mail :  zioutas@cern.ch}


\date{8 June 2004}

\maketitle

\begin{abstract}
Massive axions of the Kaluza-Klein type, created inside the solar core,
can be gravitationally trapped by the Sun itself in orbits inside/outside
the Sun, where they accumulate over cosmic times. Their spontaneous or "induced"
radiative decay can give rise to various solar phenomena, like the celebrated
solar coronal heating, which lacks a conventional explanation since its first
observation in 1939. Such and other recent observations  favour
the existence of a halo of exotic particles near the Sun. X-ray (solar)
telescopes can provide novel and important informations for astroparticle
physics and cosmology.
The underlying solar axion scenario is presented in details in ref.'s
\cite{dilella,apj607},
which can be consulted for further reading.
\end{abstract}

\section{Introduction}

In order to solve the strong CP problem, a new neutral particle with spin-parity
$0^-$, the axion, was invented (see recent ref.'s, e.g.,
\cite{raffeltg}).
Axions, along with Weakly Interacting Massive Particles (WIMPs), are the two
leading particle candidates for dark matter in the Universe. Axions
should also be  abundantly produced inside the solar core.
The expected decay to two photons ($a\rightarrow 2\gamma$) results in a lifetime
much longer than the age of the Universe.
However, in theories of extra-dimensions, the `conventional', almost massless
axions become as massive as the reaction energies involved.  In the case of the
solar axions, the expected mass spectrum of the excited Kaluza-Klein (KK) states
reaches $\sim 10$ keV/c$^2$
\cite{kk},
with a relatively short lifetime ($\tau \sim 10^{20}$s),
because of the $\tau \sim m^{-3}$ dependence.
The underlying axion-photon-photon coupling constant, $g_{a\gamma \gamma}$,
remains the same for the 'conventional' ($\approx $ massless) axion and for the
massive KK-axions.
In this approach, the KK-axions are taken as a generic example of particles
which can be created inside the hot solar core, while a small fraction of them
being highly non-relativistic ($\sim 10^{-7}$) can be gravitationally trapped
by the Sun itself in orbits  where they accumulate over cosmic times
\cite{dilella0,dilella}.
Their derived density increases enormously near the solar surface.

The spontaneous (or "induced") radiative decay of gravitationally trapped
massive axions or other particles (from the solar core) with similar properties
can give rise to a {\it self-irradiation} of the solar atmosphere
explaining the observed X-rays from the solar disk and limb;
note, an alternative conventional explanation is missing since 1939,
when Grotrian
\cite{grotrian}
made the puzzling discovery that the solar corona is $\sim 100\times$ hotter
than the underlying photosphere. Remarkably, the physical origin of the coronal
heating became "{\it one of the most fundamental problems in stellar (and solar)
astrophysics}"
\cite{guedel}.

\section{What are the Signatures ?}

According to the considered level of significance, the relevant signals of
X-rays, originating (in)directly from the decay of exotic particles around the
Sun, are :

\noindent
A)~~ {\bf The radial temperature-density profile} of the solar atmosphere near
the surface of the Sun is strikingly similar to the one of the Earth's
upper atmosphere
\cite{dilella},
which we know is being exposed to  solar illumination.
This seems to be the strongest evidence in favour of a quasi continuous
external irradiation of the Sun, providing also the energy source for  heating
the solar corona.

\noindent
B)~~ {\bf The  quiet Sun X-ray spectrum} as it has been reconstructed
from the {\it emission measure} distribution at the solar minimum
\cite{peres,dilella}
provides a high energy component up to $\sim 10$ keV.
This is the only analog solar spectrum found so far, which extends far beyond
$\sim $1-2 keV,  which is the range of interest for the axion scenario.
In fact, its energy range coincides with the Monte Carlo generated spectrum following
the example of the radiative decay of gravitationally trapped solar KK-axions,
as it has been worked out in ref.
\cite{dilella}.
In order to normalize the theoretical solar KK-axion decay X-ray spectrum to the
intensity of the reconstructed hard X-ray spectrum, the derived value of the
coupling constant axion-to-photon is :~  $g_{a\gamma \gamma} = 9.2\cdot 10^{-14} GeV^{-1}$.
Note, this value is by $\sim 4$ orders of magnitude far below the present
detection sensitivity of any known massive
axion search in underground solid state detectors
\cite{paschos}.
This demonstrates at the same time the potential power behind solar X-ray
investigations.

\noindent
Orbiting X-ray telescopes (e.g. RHESSI, or upcoming solar X-ray missions) could
measure directly the analog X-ray spectrum associated with extended quiet Sun
periods during the next solar minimum ($\sim $2005 - 2008).
In fact, the importance of X-rays from the quiet and/or not so quiet Sun
have been overlooked, inspite of the detection constraints.

\noindent C)~~ {\bf Hard X-rays from the non-flaring Sun} above
3.5 keV as it was first observed by the HXIS spectrometer on SMM
\cite{simnet}. Later, the NEAR  mission \cite{near1} provided
quiescent solar hard X-ray spectra, too. In addition, the
INTERBALL mission \cite{interball} measured hard solar X-rays,
noticing (unbiased at that time) full in the sense of the solar
axion scenario : "{\it We have found it very unexpected that there
is present quiet-Sun emission in the 10-15 keV band in the period
of the lowest solar activity (1995)}". Recently, also the RHESSI
observatory \cite{lin}, has  observed a {\it continuous} X-ray
emission, from 3 to $\sim $15 keV, with frequent (every few
minutes) microflaring \cite{rhessix1a}, when there are no
observable flares present \cite{rhessix2a}.

\noindent
The derived quiet Sun X-ray spectrum \cite{peres} below
$\sim 15$ keV corresponds rather to a $\sim$80-100 MK hot plasma
component. These photons are much more energetic than the bulk of
thermal photons from the celebrated $\sim 2$ MK solar corona
plasma (e.g. $E_{\gamma}\approx 3kT\approx 0.5$ keV). These
findings make any conventional explanation much more difficult.
Note, this part of the quiet Sun is even hotter than the $\sim$17
MK solar core. However, this is not in contradiction with the
generic KK-axion scenario, which works then, so to say, as a
built-in amplifier between the inner and the outer Sun.  This also
demonstrates the complexity of the underlying mechanism, which
must be at work to explain the solar corona X-rays.

\noindent
D)~~ {\bf The X-ray surface brightness} resulting from the decay of the trapped
axions around the Sun is expected to be continuous and quasi stable with time
and to decrease rapidly with increasing distance from the Sun
\cite{apj607}.
Possible contributions from active regions on the solar disk into the limb
region due to scattering in the X-ray telescope must be excluded.
For example, in the considered Yohkoh observations of two quiet Sun regions
\cite{apj607},
they constitute a small but not insignificant fraction
\cite{region1,region12}.

\noindent
E)~~ {\bf The inward heat flux in the solar atmosphere.}
The conventional explanation of the aforementioned two Yohkoh quiet Sun
observations
\cite{region1,region12}
suggest a mechanism that deposits {\it somehow} nonthermal energy as heat
beyond the observed range of heights  above the limb
(R$\geq 1.5$-2$\,R_{\odot}$), consistent with an inward heat flux.
In addition, within the adopted model, there is no evidence for
nonthermal heating in either the observed regions or in the inner corona
\cite{slacweb};
the paradox conclusion from this observation is : "{\it the solar wind may
supply heat to the inner corona rather than the other way around. ... The
standard view may need revision ...}"
\cite{slacweb}
\footnote{
According to the standard view, the deposition of nonthermal energy occurs in
the inner corona and this region in turn supplies heat to the upper corona and
to the solar wind, a term used to represent the continuous expansion of the
corona into planetary space
\cite{slacweb}.
}.

\noindent
F)~~ {\bf The solar} ${\bf L_x \sim B^2}$ {\bf dependence} is striking,
though it can not be considered yet as evidence, inspite of its potential
importance for  solar axion searches.
Remarkably, the relation between the observed soft X-ray flux (photon energy below
$\sim$4 keV) and the  solar magnetic flux {\bf B} can be approximated by a power
law with an averaged index close to 2, i.e., ${\bf L_x \sim B^{1.8\pm 0.4}}$,
changing smoothly over a solar cycle
\cite{bene}.
For comparison, in the ongoing CAST experiment at CERN
\cite{cast}
we expect an axion-to-photon conversion probability
\cite{zioutas}
inside the transverse magnetic field to depend also  on  ${\bf B^2}$.
Therefore, it is interesting to explain quantitatively whether the CAST working
principle may well take place, for some reason(s), more efficiently near the
Sun
\cite{hoffmann}.
If this happens to be the case, it might be suggestive for a new experimental
approach in axion research 1 AU from the Sun.

\noindent
For the invented massive solar KK-axions
\cite{dilella},
the coherent conversion probability inside a magnetic field must be suppressed,
at least at first sight, due to the $(m_{axion})^{-2}$ dependence of the
coherence length.
But, if finally the observed  ${\bf L_x \sim B^2}$ relation reflects an
electromagnetically induced axion materialization (whatever the physical
mechanism is found to be behind it at the end), this might open a window to the
11-years solar cycle, because the solar X-ray variability during the solar
cycle is strong and well established.

\noindent
In addition, such magnetic field related effects could provide also the
explanation for a plethora of local solar X-ray emission phenomena. If this is
axion related, it is an important perspective.

\section{Conclusions}

A short summary of potential solar signals are given, which are
consistently in favour of the generic KK-axion scenario, thus extending
the lines of reasoning given in details in the ref.'s
\cite{dilella,apj607}.
The basic idea behind the  axion scenario is the following :
massive axion(-like) particles of the Kaluza-Klein type are created inside the
solar core. Some of them are highly non-relativistic and they can be
gravitationally trapped by the Sun itself, in orbits inside and outside of it,
where they accumulate over cosmic time scales.
Such highly dense  exotic particles around the Sun can decay
(spontaneously and/or "induced"), explaining thus  various, yet as
mysterious characterised,  solar phenomena.

The solar axion scenario  provides then the required continuous and steady power
input into the solar atmosphere. The photon energies are rather high
for (quiet) solar standards, for which, so far,  any conventional
explanation is missing.  However, depending on the local physical
conditions, e.g. magnetic field strength, plasma density, etc., an additional
axion-to-photon conversion mechanism may enhance locally this energy input,
providing an explanation also for the not so rare local solar X-ray
effects.

Therefore, the ongoing as well as the upcoming solar X-ray spectroscopy
missions, along with additional theoretical work about possible, yet unforeseen,
dynamical behaviour of the highly dense solar axion(-like) cloud, could provide
mutual feedback. This might prove to be essential for a deeper
understanding of those relatively high energy solar phenomena.
The same axion scenario should also apply to other places in the Universe,
like the Galactic Center and the X-ray luminous Clusters of Galaxies
\cite{dilella},
which, in this context, are also dominated by interesting "mysterious"
observations
\cite{preparation}.

\newpage
\noindent {\bf Acknowledgment}

\noindent
I would like to thank the organisers of this conference for their kind
invitation and hospitality. Marie Di Marco is acknowledged  for a carefull
reading of the manuscript.


\newpage


\begin{thebibliography}{99}

\bibitem{raffeltg} G.G. Raffelt, Phys. Rep. {\bf 333} (2000) 593,
 Ann. Rev. Nucl. Part. Sci. {\bf 49} (1999) 163;
 R. Bradley et al., Rev. Mod. Phys. {\bf 75} (2003) 777.

\bibitem{kk} K.R. Dienes, E. Dudas, T. Gherghetta, Phys. Rev. {\bf D62}
 (2000) 105023;

 L. DiLella, A. Pilaftsis, G. Raffelt, K. Zioutas, Phys. Rev. {\bf D62} (2000)
 125011.

\bibitem{dilella0}  L. DiLella, K. Zioutas, Phys. Lett.   {\bf B531} (2002) 175.

\bibitem{dilella} L. DiLella, K. Zioutas, Astroparticle Physics
  {\bf 19} (2003) 145.

\bibitem{grotrian} W. Grotrian, Naturwissenschaften {\bf 27} (1939) 214.

\bibitem{guedel} M. G\"udel, M. Audard, V.L. Kashyap, J.J.Drake, E.F.Guinan, ApJ. {\bf 582} (2003) 423.

\bibitem{peres} G. Peres, S. Orlando, F. Reale, R. Rosner, H. Hudson,
 ApJ. {\bf 528} (2000) 537.

\bibitem{paschos} E.A. Paschos, K. Zioutas, Phys. Lett. {\bf B323} (1994) 367.

\bibitem{simnet} G.M. Simnett, {\it in} The Many Faces of the Sun. A Summary of
 the Results from NASA's SMM, edts. K.T. Strong et al., Springer
 Verlag, Berlin (1998) 201.

\bibitem{near1} J.I. Trombka et al., J. Geophys. Res. {\bf 102} ($\#$E10) (1997)
 23729 (Fig. 20), and, Science {\bf 289} (2000) 2101 (Fig. 2c);
 R. Starr et al., ICARUS {\bf 147} (2000) 498 (Fig. 11).

\bibitem{interball} M. Siarkowski, J. Sylwester, S. Gburek, Z. Kordylewski,
 {\bf ESA SP-448}, ed. A. Wilson (1999) p.176;
 J. Sylwester et al., Solar Phys. {\bf 197} (2000) 337 (see Fig. 3).

\bibitem{lin} G.J. Hurford, S. Krucker, AAS 200th meeting, Albuquerque, NM, June
 2002, Session [76.04]; Bulletin of the AAS {\bf 34} (2002);
 R.P. Lin et al., Solar Physics {\bf 210} (2002) 3.


\bibitem{rhessix1a} R. Lin,  Abstract Nr. 532, in
 http://www.astronomy2003.com/ABSTRACT$\_$BOOK.pdf, p. 68, and,
 Adv. Space Res. {\bf 32} (2003) 1001.

\bibitem{rhessix2a} J.M. McTierman, J.A. Klimchuk, 34th Solar Division Meeting,
 June 2003, Session [18.08], in Bulletin of the AAS,  35 ($\#$3) (2003).

\bibitem{apj607} K. Zioutas, K. Dennerl, L. DiLella, D.H.H. Hoffmann, J. Jacoby,
 Th. Papaevangelou,  ApJ. {\bf 607} (2004) 575.

\bibitem{region1} P.A. Sturrock, M.S. Wheatland, L.W. Acton,
 ApJ. {\bf 461} (1996) L115.

\bibitem{region12} M.S. Wheatland, P.A. Sturrock, L.W. Acton,
  ApJ. {\bf 482} (1997) 510.

\bibitem{slacweb} http://www.stanford.edu/dept/physics/newsletter/96/corona.html

\bibitem{bene} E.E. Benevolenskaya, A.G. Kosovichev, J.R. Lemen, P.H. Scherer,
 G.L. Slater,

 ApJ. {\bf 571} (2002) L181.

\bibitem{cast} Th. Dafni, these proceedings, see also http://www.cern.ch/CAST

\bibitem{zioutas} K. Zioutas et al., Nucl. Instr. Meth. in Phys. Res.
 {\bf 425} (1999) 482 (astro-ph/9801176).

\bibitem{hoffmann} D.H.H. Hoffmann and K. Zioutas, {\it in preparation.}

\bibitem{preparation}  {\it In preparation}.

\end{thebibliography}
\end{document}